\documentstyle[12pt]{article}
\input psfig.sty
\def\nn{\noindent}
\def\Re{{\cal R \mskip-4mu \lower.1ex \hbox{\it e}\,}}
\def\Im{{\cal I \mskip-5mu \lower.1ex \hbox{\it m}\,}}
\def\ie{{\it i.e.}}
\def\eg{{\it e.g.}}
\def\etc{{\it etc}}
\def\etal{{\it et al.}}

\def\sub#1{_{\lower.25ex\hbox{$\scriptstyle#1$}}}
\def\tev{\,{\ifmmode\mathrm {TeV}\else TeV\fi}}
\def\gev{\,{\ifmmode\mathrm {GeV}\else GeV\fi}}
\def\mev{\,{\ifmmode\mathrm {MeV}\else MeV\fi}}
\def\to{\rightarrow}

\def\subw{_{\rm w}}
\def\mh{\ifmmode m\sbl H \else $m\sbl H$\fi}
\def\mch{\ifmmode m_{H^\pm} \else $m_{H^\pm}$\fi}
\def\mt{\ifmmode m_t\else $m_t$\fi}
\def\mc{\ifmmode m_c\else $m_c$\fi}
\def\mz{\ifmmode M_Z\else $M_Z$\fi}
\def\mw{\ifmmode M_W\else $M_W$\fi}
\def\mws{\ifmmode M_W^2 \else $M_W^2$\fi}
\def\mhs{\ifmmode m_H^2 \else $m_H^2$\fi}   
\def\mzs{\ifmmode M_Z^2 \else $M_Z^2$\fi}
\def\mts{\ifmmode m_t^2 \else $m_t^2$\fi}
\def\mcs{\ifmmode m_c^2 \else $m_c^2$\fi}
\def\mchs{\ifmmode m_{H^\pm}^2 \else $m_{H^\pm}^2$\fi}
\def\ztwo{\ifmmode Z_2\else $Z_2$\fi}
\def\zone{\ifmmode Z_1\else $Z_1$\fi}
\def\mtwo{\ifmmode M_2\else $M_2$\fi}
\def\mone{\ifmmode M_1\else $M_1$\fi}
\def\tb{\ifmmode \tan\beta \else $\tan\beta$\fi}
\def\xw{\ifmmode x\subw\else $x\subw$\fi}
\def\ch{\ifmmode H^\pm \else $H^\pm$\fi}
\def\lum{\ifmmode {\cal L}\else ${\cal L}$\fi}
\def\inpb{\,{\ifmmode {\mathrm {pb}}^{-1}\else ${\mathrm {pb}}^{-1}$\fi}}
\def\infb{\,{\ifmmode {\mathrm {fb}}^{-1}\else ${\mathrm {fb}}^{-1}$\fi}}
\def\epem{\ifmmode e^+e^-\else $e^+e^-$\fi}
\def\ppb{\ifmmode \bar pp\else $\bar pp$\fi}
\def\bsg{\ifmmode B\to X_s\gamma\else $B\to X_s\gamma$\fi}
\def\bsll{\ifmmode B\to X_s\ell^+\ell^-\else $B\to X_s\ell^+\ell^-$\fi}
\def\bstt{\ifmmode B\to X_s\tau^+\tau^-\else $B\to X_s\tau^+\tau^-$\fi}
\def\lamt{\ifmmode \tilde\lambda\else $\tilde\lambda$\fi}
\def\shat{\ifmmode \hat s\else $\hat s$\fi}
\def\that{\ifmmode \hat t\else $\hat t$\fi}
\def\uhat{\ifmmode \hat u\else $\hat u$\fi}

\newskip\zatskip \zatskip=0pt plus0pt minus0pt
\def\matth{\mathsurround=0pt}

\def\atversim#1#2{\lower0.7ex\vbox{\baselineskip\zatskip\lineskip\zatskip
  \lineskiplimit 0pt\ialign{$\matth#1\hfil##\hfil$\crcr#2\crcr\sim\crcr}}}

\renewcommand{\thefootnote}{\fnsymbol{footnote}}

\begin{document} \begin{titlepage} 
\rightline{\vbox{\halign{&#\hfil\cr
%&DRAFT\cr
&SLAC-PUB-8114\cr
&April 1999\cr}}}
\begin{center}

{\Large\bf Tests of Low Scale Gravity via Gauge Boson Pair Production in 
$\gamma \gamma$ Collisions}
\footnote{Work supported by the Department of 
Energy, Contract DE-AC03-76SF00515}
\medskip

\normalsize 
{\large Thomas G. Rizzo } \\
\vskip .3cm
Stanford Linear Accelerator Center \\
Stanford University \\
Stanford CA 94309, USA\\
\vskip .3cm

\end{center}

\begin{abstract} 
Arkani-Hamed, Dimopoulos and Dvali have recently proposed that gravity may 
become strong at energies near 1 TeV thus removing the hierarchy problem. 
This scenario can be tested in several ways at present and future colliders. 
In this paper we examine the exchange of towers of Kaluza-Klein gravitons 
and their influence on the production of pairs of massive gauge bosons in 
$\gamma \gamma$ collisions. These tower exchanges are shown to lead to a 
new dimension-8 operator that can significant alter the Standard Model 
expectations for these processes. The role of polarization for both the 
initial state photons and the final state gauge bosons in improving 
sensitivity to graviton exchange is emphasized. We find that the discovery 
reach for graviton tower exchange in the $\gamma \gamma \to W^+W^-$ channel 
to be significantly greater than for any other process so far examined.
\end{abstract} 

%\vskip0.45in
%\begin{center}

%Submitted to Physical Review {\bf D}.

%\end{center}

\renewcommand{\thefootnote}{\arabic{footnote}} \end{titlepage}

%%%%%%%%%%%%%%%%%%%%%%%%%%%%%%%---- Put text here

\section{Introduction}

Arkani-Hamed, Dimopoulos and Dvali(ADD)~{\cite {nima}} have recently 
proposed a uniquely interesting solution to the hierarchy problem. ADD 
hypothesize the existence of $n$ additional large spatial dimensions in 
which gravity (and perhaps Standard Model singlet fields) can live, called 
`the bulk', whereas all of the fields of the Standard Model(SM) are 
constrained to lie on `the wall', which is our 
conventional 4-dimensional world. Gravity thus appears to us as weak in 
ordinary 4-dimensional space-time since we merely observe it's action on the 
wall. It has been shown~{\cite {nima}} that such a scenario can emerge in 
string models where the effective Planck scale in the bulk is identified 
with the string scale. In such a theory the hierarchy is 
removed by postulating that the string or effective Planck scale in 
the bulk, $M_s$, is not far above the weak scale, \eg, a few TeV. Gauss' Law 
then provides a link between the values of $M_s$, the conventional 
Planck scale $M_{pl}$, and the size of the compactified extra dimensions, $R$, 
\begin{equation}
M_{pl}^2 \sim R^nM_s^{n+2}\,,
\end{equation}
where the constant of proportionality depends not only on the value of $n$ 
but upon the geometry of the compactified dimensions. If $M_s$ 
is near a TeV then $R\sim 10^{30/n-19}$ meters; for separations between two 
masses less than $R$ the gravitational force law becomes $1/r^{2+n}$. 
For $n=1$, $R\sim 10^{11}$ meters and is thus obviously excluded, but, 
for $n=2$ one obtains $R \sim 1$~mm, which is at the edge of the sensitivity 
for existing experiments{\cite {test}}. For $2<n$, 
the value of $R$ is only further reduced and thus we 
may conclude that the range $2\leq n$ is of phenomenological interest. 
Astrophysical arguments based on supernova cooling{\cite {astro}} seem to 
require that 
$M_s>50$ TeV for $n=2$, but allow $M_s\sim 1$ TeV for $n>2$ while cosmological 
arguments suggest{\cite {lhall} an even stronger constraint, $M_s>110$ TeV, 
for $n=2$.

The Feynman rules for this scenario are obtained by considering 
a linearized theory of gravity 
in the bulk, decomposing it into the more familiar 4-dimensional states and 
recalling the existence of Kaluza-Klein towers for each of the conventionally 
massless fields. The entire set of fields in the K-K tower couples in an 
identical fashion to the particles of the SM. By considering the forms of the 
$4+n$  
symmetric conserved stress-energy tensor for the various SM fields and by 
remembering that such fields live only on the wall one may derive all of the 
necessary couplings. An important result of 
these considerations is that only the massive spin-2 K-K towers (which couple 
to the 4-dimensional stress-energy tensor, $T^{\mu\nu}$) and spin-0 K-K 
towers (which couple proportional to the trace of $T^{\mu\nu}$) are of 
phenomenological relevance as all the spin-1 fields can be shown to decouple 
from the particles of the SM. For processes that involve massless fields at 
at least one vertex, as will be the case below, the contributions of 
the spin-0 fields can also be ignored.

The detailed phenomenology of the ADD model has begun to be explored for a 
wide ranging set of processes in 
a rapidly growing series of recent papers~{\cite {pheno}}. 
Given the Feynman rules as developed by Guidice, Rattazzi and Wells and by 
Han, Lykken and Zhang{\cite {pheno}}, it appears that the 
ADD scenario has two basic classes of collider tests. In the first class, 
type-$i$, 
a K-K tower of gravitons can be emitted during a decay or scattering process 
leading to a final state with missing energy. The rate for such processes is 
strongly dependent on the number of extra dimensions as well as the exact 
value of $M_s$. However, in this case the value of $M_s$ is directly probed. 
In the second class, type-$ii$, 
which we consider here, the exchange of a K-K graviton tower between SM or 
MSSM fields can lead to almost 
$n$-independent modifications to conventional cross sections and 
distributions or they can possibly lead to new interactions. The exchange of 
the graviton K-K tower leads to a set of effective color and flavor singlet 
contact interaction operator of dimension-eight with the overall scale set by 
the cut-off in the tower summation, $\Lambda$. Naively $\Lambda$ and $M_s$ 
should be of comparable magnitude and so one introduces a 
universal overall order one coefficient for these 
operators, $\lambda$ ( whose value is unknown but can be approximated by a 
constant which has conventionally been 
set to $\pm 1$) with $\Lambda$ being replaced by $M_s$. The fact that 
$\Lambda$ can be smaller than $M_s$ is thus particularly important when 
considering the case of 2 extra dimensions due to the rather strong 
astrophysical and cosmological constraints that then apply. 
We note that $\lambda$ 
can in principle have either sign since the unknown physics above the cut-off 
can make an additional universal 
contribution to the coefficient of the relevant 
amplitude of indeterminate sign. Given the kinematic 
structure of these operators the modifications in the relevant 
cross sections and distributions can be directly calculated. 

In what follows we will consider the production of massive gauge boson pairs in 
$\gamma \gamma$ collisions via the exchange of a K-K tower of gravitons (the 
process $\gamma\gamma \to \gamma \gamma$ having been 
considered{\cite {pheno}} elsewhere). 
Such reactions can be examined in detail at future linear colliders via the 
Compton back-scattering of laser photons off of high energy colliding 
$e^+e^-$ or $e^-e^-$ beams{\cite {telnov}}. As we note below, the role of 
polarization, in both the initial state as well as the final state, is 
crucial in separating the graviton signal from the SM background and extending 
the search reach. We then will demonstrate that the discovery 
reach for graviton tower exchange in the $WW$ channel is greater than any other 
process so far examined. The corresponding reach in the $ZZ$ case will be 
shown to be rather modest and comparable to that found for the 
$\gamma \gamma \to \gamma \gamma$ process. We also note the added bonus 
feature associated with $\gamma\gamma$ collisions. Whereas, \eg , 
$e^+e^-\to f\bar f$ can receive contributions from $Z$ and $\gamma$ 
Kaluza-Klein towers in some models, the processes $\gamma \gamma \to X\bar X$ 
do not at tree level and are thus can provide very clean signatures for 
graviton exchange.

\section{$\gamma \gamma \to VV$ via Graviton Exchange}

$\gamma \gamma$ collisions offer a unique and distinct window on the 
possibility of new physics in a particularly clean environment. At tree 
level the cross section for particle pair production, if it is allowed by the 
gauge symmetries, 
depends only gauge couplings. Unlike gauge boson pair production in 
$e^+e^-$ collisions, however, $P$, $C$, plus the Bose symmetry of the 
initial state photons forbids the existence of non-zero values, at the 
tree level, for either forward-backward angular asymmetries or left-right 
forward-backward polarization asymmetries. These were both found to be 
powerful tools in probing for K-K graviton tower exchanges in 
the $e^+e^-$ initiated channels{\cite {pheno}}. 
In the case of $\gamma\gamma$ collisions our 
remaining tools are the angular distributions of the produced pairs of 
vector bosons, their resulting states of polarization, and the cross section's 
sensitivity to the polarization of the initial state photons. 

The exchange of a tower of K-K gravitons leads to a new tree level 
contribution to the process 
$\gamma(k_1,\epsilon_1)\gamma(k_2,\epsilon_2)\to 
V(p_1,V_1)V(p_2,V_2)$, where $k_{1,2}$ are incoming and $p_{1,2}$ are 
outgoing momenta and $\epsilon_{1,2}(V_{1,2})$ represent the polarization 
vectors for the photons(vector bosons), respectively. Following the Feynman 
rules of either Guidice, Rattazzi and Wells or Han, Lykken 
and Zhang{\cite {pheno}} we can immediately write down the relevant matrix 
element which has the form of a dimension-8 operator: 

\begin{equation}
{\cal M}_{grav}={8\lambda K\over M_s^4}\left[M_1+M_2+M_3+M_4+M_5\right]\,,
\end{equation}
where
\begin{eqnarray}
M_1 &=& {s^2\over {2}}\bigg[\epsilon_1\cdot V_1^* \epsilon_2\cdot V_2^*+
\epsilon_1\cdot V_2^* \epsilon_2\cdot V_1^* \bigg]+s\epsilon_1 \cdot 
\epsilon_2\bigg[k_1\cdot V_1^* k_2\cdot V_2^*+k_1\cdot V_2^* k_2\cdot 
V_1^*\nonumber \\
&-& k_1\cdot k_2 V_1^* \cdot V_2^* \bigg] \nonumber \\
M_2 &=& sV_1^* \cdot V_2^* \bigg[p_1\cdot \epsilon_1 p_2\cdot \epsilon_2+
p_1\cdot \epsilon_2 p_2\cdot \epsilon_1 \bigg]+ 2\epsilon_1 \cdot \epsilon_2 
V_1^* \cdot V_2^* \bigg[p_1 \cdot k_1 p_2 \cdot k_2+p_1 \cdot k_2p_2 \cdot 
k_1\nonumber \\
&-& p_1 \cdot p_2 k_1 \cdot k_2\bigg]\nonumber \\
M_3 &=& -sp_1\cdot V_2^*\bigg[\epsilon_1 \cdot V_1^* \epsilon_2 \cdot p_2
+\epsilon_2 \cdot V_1^* \epsilon_1 \cdot p_2\bigg]-sp_2\cdot V_1^*
\bigg[\epsilon_1 \cdot V_2^* \epsilon_2 \cdot p_1+
\epsilon_2 \cdot V_2^* \epsilon_1 \cdot p_1\bigg]\nonumber \\
M_4 &=& -2p_1\cdot V_2^* \epsilon_1 \cdot \epsilon_2 \bigg[k_1\cdot V_1^* 
k_2\cdot p_2+k_2\cdot V_1^* k_1\cdot p_2-p_2\cdot V_1^* k_1\cdot k_2\bigg]\\ 
M_5 &=& -2p_2\cdot V_2^* \epsilon_1 \cdot \epsilon_2 \bigg[k_1\cdot V_2^* 
k_2\cdot p_1+k_2\cdot V_2^* k_1\cdot p_1-p_1\cdot V_2^* k_1
\cdot k_2\bigg]\nonumber \,,
\end{eqnarray}
with $s$ being the usual Mandelstam variable. We note that in the expression 
above $M_s$ should actually be the cut-off scale used in performing the 
summation over the K-K tower in the $s$-channel. In principle, as discussed 
above, these two scales may differ by by some factor of 
order unity which we thus incorporate into the parameter $\lambda$. 
For $K=1(\pi/2)$ we 
recover the normalization convention employed by Hewett(Guidice, Rattazzi and 
Wells){\cite {pheno}}; we will take $K=1$ in the numerical analysis that 
follows but keep the factor in our analytical expressions. 
We recall from the original Hewett analysis that $\lambda$ is to be treated as 
a parameter of order unity whose sign is undetermined and that, given the 
scaling relationship between $\lambda$ and $M_s$, experiments in the 
case of processes 
of type-$ii$ actually probe only the combination $M_s/|K\lambda|^{1/4}$. 
For simplicity in what follows we will numerically set $|\lambda|=1$ and 
employ $K=1$ but we caution the reader about this 
technicality and quote our sensitivity to $M_s$ for $\lambda=\pm 1$ as is now 
the standard tradition.

In the center of mass frame we imagine that the initial photons are coming in 
along the $z$-axis with the outgoing vector bosons making an 
angle $\theta$ with 
respect to this axis. Using $\beta^2=1-4m^2/s$, $m$ being the mass of the field 
$V$, and recalling that the transverse initial state photons are assumed to be 
circularly polarized, we can describe the complete kinematics for the process 
in terms of the following four-vectors:
\begin{equation}
\begin{array}{rcl}
k_{1\mu} &=& {{\sqrt s}\over {2}}(1,0,0,1)\\
p_{1\mu} &=& {{\sqrt s}\over {2}}(1,\beta s_\theta,0,\beta c_\theta)\\ 
\epsilon_1^\mu &=& -{1\over {\sqrt 2}}(0,\lambda_1,i,0)\\
V_{1L}^{\mu *} &=& {{\sqrt s}\over {2m}}(-\beta,s_\theta,0,c_\theta)\\
V_{1T}^{\mu *} &=& {1\over {\sqrt 2}}(0,-\lambda_1^Vc_\theta,i,\lambda_1^V
s_\theta)
\end{array}\qquad
\begin{array}{rcl}
k_{2\mu} &=& {{\sqrt s}\over {2}}(1,0,0,-1)\\
p_{2\mu} &=& {{\sqrt s}\over {2}}(1,-\beta s_\theta,0,-\beta c_\theta)\\ 
\epsilon_2^\mu &=& {1\over {\sqrt 2}}(0,-\lambda_2,i,0)\\
V_{2L}^{\mu *} &=& -{{\sqrt s}\over {2m}}(\beta,s_\theta,0,c_\theta)\\
V_{2T}^{\mu *} &=& {1\over {\sqrt 2}}(0,-\lambda_2^Vc_\theta,-i,\lambda_2^V
s_\theta)
\end{array}
\end{equation}
with $s_\theta=\sin \theta$ and $z=c_\theta=\cos \theta$. The use of covariant 
and contravariant indices in these expressions should be noted by the reader.
Here the indices $T,L$ on the polarization vectors $V_{1,2}$ denote states 
of transverse and longitudinal polarization. The quantities $\lambda_{1,2}$ 
and $\lambda_{1,2}^V$ parameterize the two transverse polarization states 
of the photons and vector bosons, respectively, and 
take on the values $\pm 1$. Given these kinematical expressions we can 
immediately evaluate all of the dot products appearing in the matrix element 
for any given choice of photon and/or $V$ polarizations. The resulting 
differential cross section for any particularly chosen set of helicities 
can then be written as 
\begin{equation}
{d\sigma \over {dz}}={1\over {1+\delta_{VZ}}}{\beta \over {32\pi s}}
|{\cal M}_{SM}+{\cal M}_{grav}|^2\,,
\end{equation}
where ${\cal M}_{SM}$ symbolically represents the more conventional 
contribution to these matrix elements arising in the SM or the MSSM, \etc. Note 
the Kronecker delta factor in the denominator in the case of the identical 
particle $ZZ$ final state.

Unfortunately $\gamma\gamma$ collisions at future linear colliders will not 
be as straightforward as the description above would indicate since the 
photons in the collision will not be either monoenergetic or in a unique 
state of polarization. 
Polarized $\gamma\gamma$ collisions may be possible at future $e^+e^-$ 
colliders through the use of Compton backscattering of polarized low energy 
laser beams off of polarized high energy electrons{\cite {telnov}}. The 
resulting 
backscattered photon distribution, $f_\gamma(x=E_\gamma/E_e)$, is as stated 
above far from 
monoenergetic and is cut off above $x_{max}\simeq 0.83$ implying that 
the colliding photons are significantly softer than the parent lepton beam 
energy. As one sees, this cutoff at 
large $x$, $x_{max}$, implies that the $\gamma \gamma$ center of mass energy 
never exceeds $\simeq 0.83$ of the parent collider and this has resulted in a 
significantly degraded $M_s$ sensitivity for final states involving fermion or 
scalar pairs{\cite {pheno}}. In 
addition, both the shape of the function $f_\gamma$ and the average helicity 
of the produced $\gamma$'s are quite sensitive to the 
polarization state of both the initial laser ($P_l$) and electron ($P_e$) 
whose values fix the specific distribution. 

While it is anticipated that the 
initial laser polarization will be near $100\%$, \ie, $|P_l|=1$, the electron 
beam polarization is expected to be be near $90\%$, \ie, $|P_e|=0.9$. We will 
assume these values in the analysis that follows. With two photon `beams' 
and the choices $P_l=\pm 1$ and $P_e=\pm 0.9$ to be made for each beam it 
would appear that 16 distinct polarization-dependent cross sections need to be 
examined. However, due to the exchange symmetry between the two photons and the 
fact that a simultaneous flip in the signs of all the polarizations leaves the 
product of the fluxes, mean helicities and the cross sections 
invariant, we find that there 
only six physically distinct initial polarization 
combinations{\cite {chak}}. In what follows we will label these 
possibilities by the corresponding signs of the electron and laser 
polarizations as $(P_{e1},P_{l1},P_{e2},P_{l2})$, For example, the 
configuration $(-++-)$ corresponds to $P_{e1}=-0.9$, $P_{l1}=+1$, $P_{e2}=0.9$ 
and $P_{l2}=-1$. Clearly some of these polarization 
combinations will be more sensitive to the effects of K-K towers of gravitons 
than will others so our analysis can be used to pick out those particular 
cases. Within this framework we must view the above differential cross section 
as that of a partonic subprocess in analogy with hadronic collisions, \ie, 
$d \sigma \to d \hat \sigma$  and we identify $s \to \hat s=s_{e^+e^-}x_1x_2$.

For any given choice of the initial state laser and electron polarizations, 
labelled by $(a,b)$ below, we can immediately write down the appropriate 
cross section by folding in the corresponding photon fluxes and integrating:
\begin{equation}
{d\sigma^{ab}\over {dz}}=\int ~dx_1~\int~dx_2~f^a_\gamma(x_1,\xi_1)
f^b_\gamma(x_2,\xi_2)\Bigg[{{1+\xi_1\xi_2}\over {2}}
{d\hat \sigma_{++} \over {dz}}+{{1-\xi_1\xi_2}\over {2}}
{d\hat \sigma_{+-} \over {dz}}\Bigg]\,,
\end{equation}
where we explicitly note the dependence of the fluxes on the mean helicities 
$\xi_{1,2}$ which are known functions of energy as well as both the initial 
state laser and electron beam polarization. The $++$ and $+-$ labels on the 
subprocess cross sections indicate the appropriate values of $\lambda_{1,2}$ 
to chose in their evaluation. In order to obtain ${d\sigma \over {dz}}$ the 
polarizations of the vector bosons in the final state must be either 
specified or summed over. Similarly, we can obtain the unpolarized cross 
section by averaging over the initial state photon polarizations.
Given the fluxes{\cite {telnov}} these integrals are easily evaluated 
numerically. The upper limit of both integrals is just $x_{max}$. 
In the present case the kinematics require the photon energies to satisfy the 
constraint $\tau=\hat s/s=x_1x_2\geq 4m^2/s=\tau_{min}$ which, together with 
the value of $x_{max}$, then determines the lower bounds on both $x_{1,2}$: 
$x_1^{min}=\tau_{min}/x_{max}$ and $x_2^{min}=\tau_{min}/x_1$.

\begin{figure}[htbp]
\centerline{
\psfig{figure=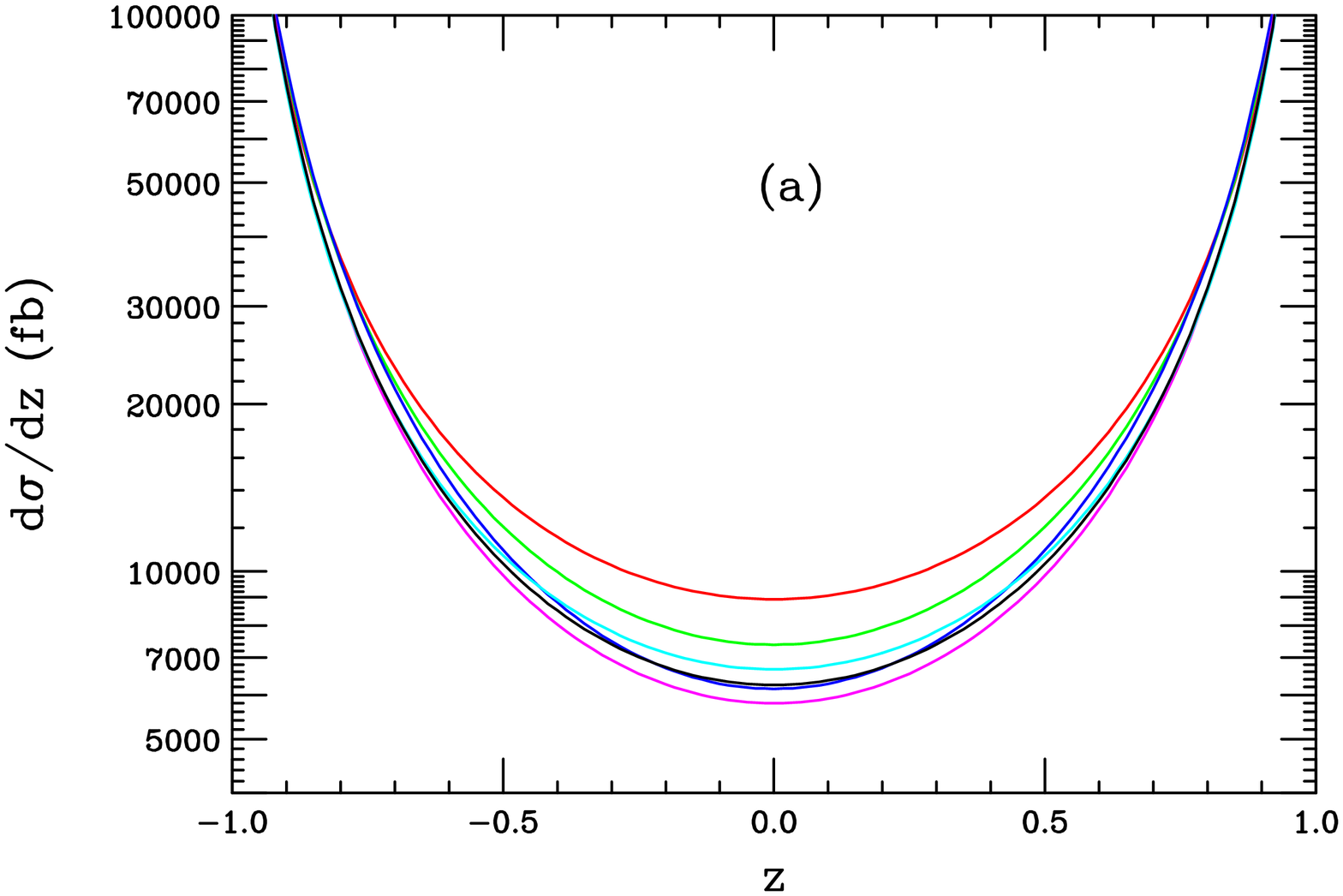,height=8.9cm,width=9.1cm,angle=90}
\hspace*{5mm}
\psfig{figure=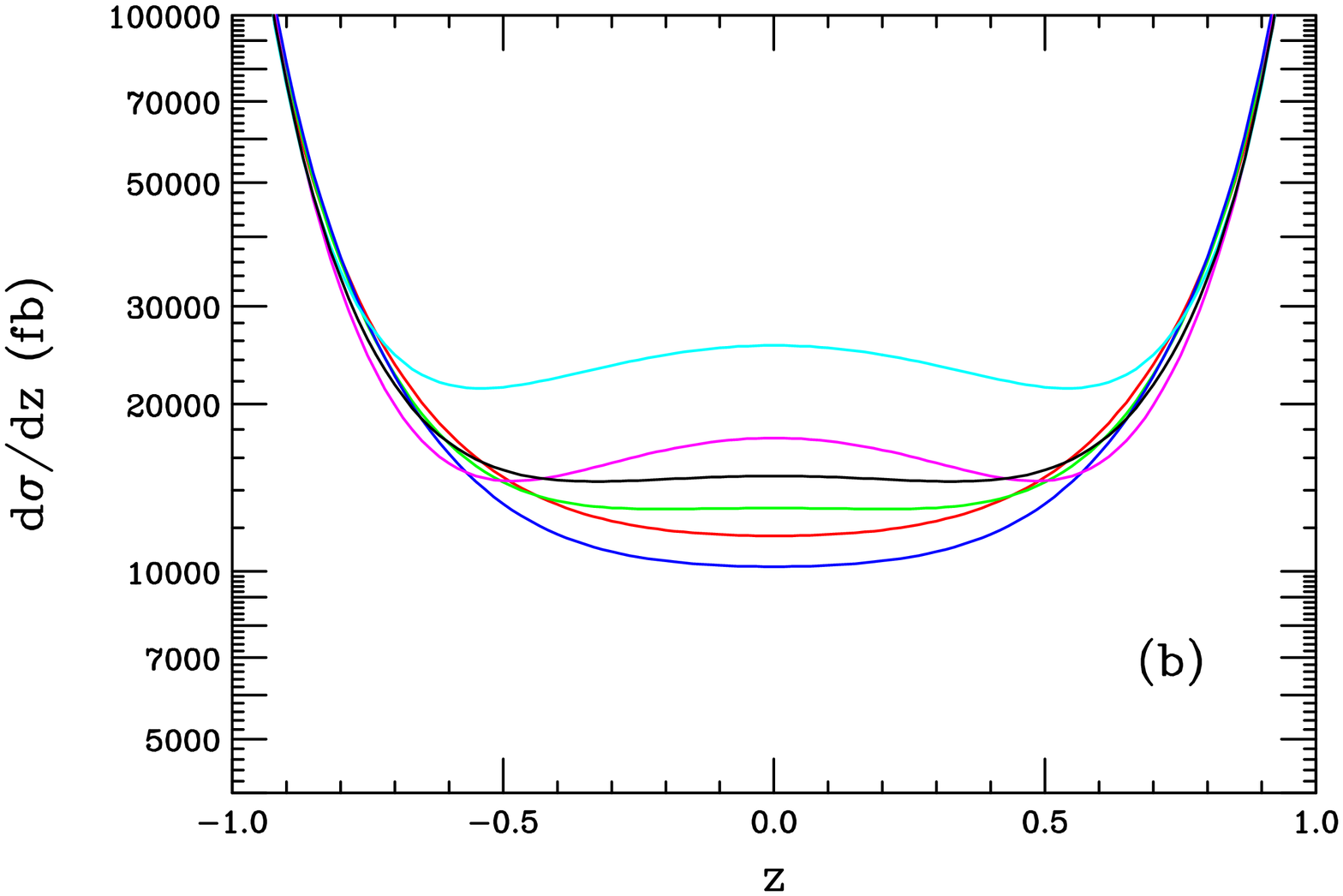,height=8.9cm,width=9.1cm,angle=90}}
\vspace*{0.15cm}
\centerline{
\psfig{figure=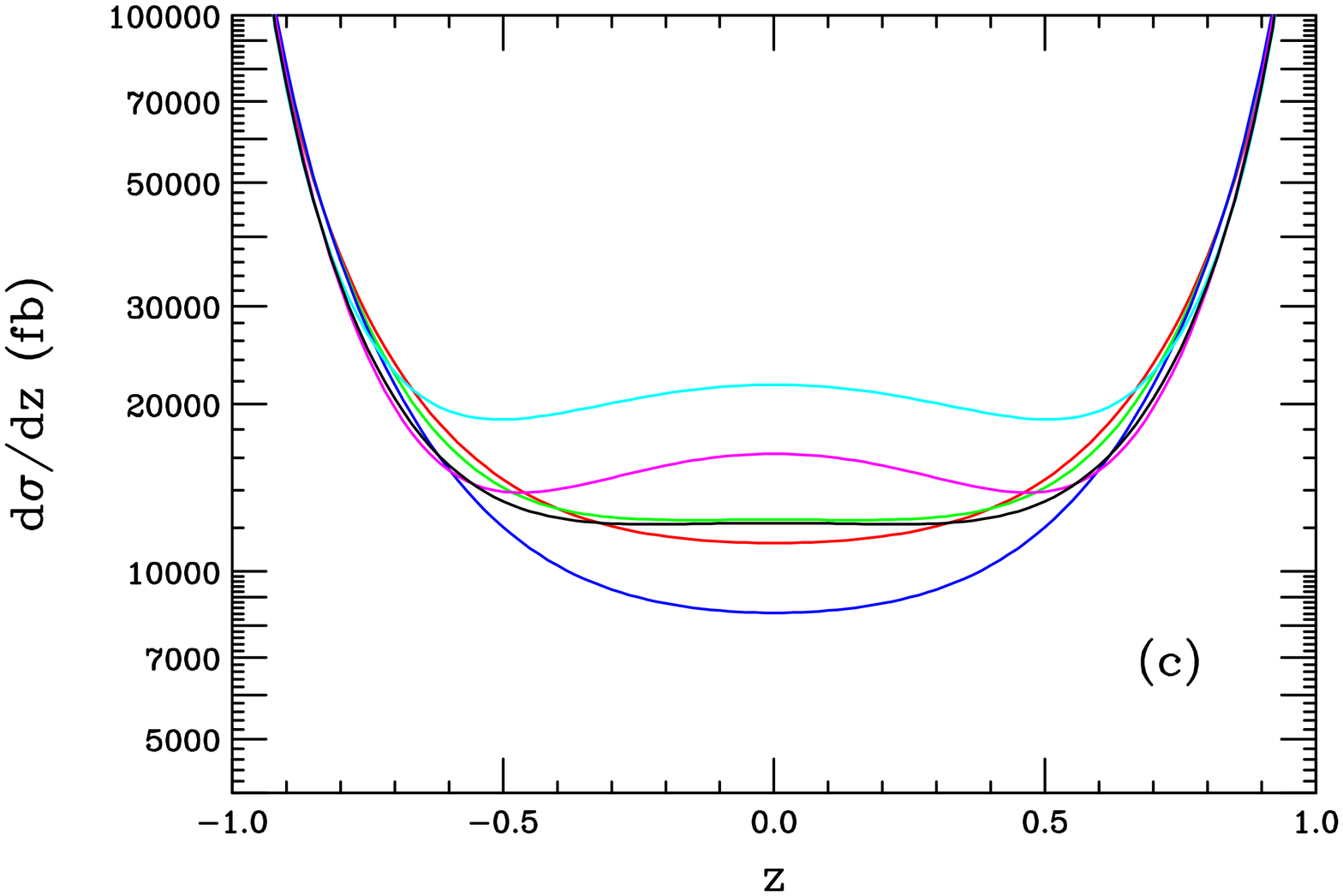,height=8.9cm,width=9.1cm,angle=90}}
\vspace*{0.05cm}
\caption{Differential cross section for $\gamma\gamma \to W^+W^-$ at a 1 TeV 
$e^+e^-$ collider for (a)the SM and with $M_s=2.5$ TeV with (b)$\lambda=1$ or 
(c)$\lambda=-1$. In (a) from top to bottom in the center of the figure 
the helicities are $(++++)$, 
$(+++-)$, $(-++-)$, $(++--)$, $(+---)$, and $(+-+-)$; in (b) they are  
$(-++-)$, $(+-+-)$, $(+++-)$, $(+---)$, $(++++)$, and $(++--)$;  in (c) they 
are  $(-++-)$, $(+-+-)$, $(+---)$, $(+++-)$, $(++++)$, and $(++--)$.}
\label{fig1}
\end{figure}

\section{Results}

Now that we have specified all of the relevant machinery we can now turn to 
some results. We consider the two cases $V=W$ and $V=Z$ separately.

\subsection{$\gamma \gamma \to W^+W^-$}

The tree-level SM helicity amplitudes for the process 
$\gamma \gamma \to W^+W^-$ have been known for some time{\cite {wsm}} and even 
the complete one-loop corrections within the SM are also known{\cite {loop1}. 
In what follows 
we will ignore these loop corrections but remind the reader that they must be 
employed in a complete analysis that also includes detector effects \etc. 
Given these 
amplitudes, we can immediately calculate the relevant differential cross 
section combining SM and graviton tower exchanges. To be definitive we will 
assume $\sqrt s=\sqrt s_{e^+e^-}$=1 TeV with $M_s$=2.5 TeV for purposes of 
demonstration. 

The results are shown in Fig.\ref{fig1} for the SM case as 
well as when the K-K tower is turned on with either sign of $\lambda$ for all 
six initial helicity combinations. In the SM case the shape of the angular 
distribution is easily understood by recalling that the 
$\gamma \gamma \to W^+W^-$ reaction takes place through $t$- and $u$-channel 
$W$ exchanges as well as a $\gamma \gamma W^+W^-$ four-point interaction. The 
$t$- and $u$-channel exchanges thus lead to a sharply rising cross section 
in both the forward and backward directions.  Note that in the SM there is no 
dramatically strong sensitivity to the initial state lepton and laser 
polarizations in this case and all of the curves have roughly the same shape. 
In all cases the total cross section, even after generous angular cuts, is 
quite enormous, of order $~\sim 100$ pb, providing huge statistics to look 
for new physics influences.  
When the graviton terms are turned on there are several effects. First, all 
of differential cross section distributions become somewhat more shallow, 
particularly in the case 
of $\lambda=1$, but there is little change in the forward and backward 
directions due to the dominance of the SM poles. Second, there is now a 
clear and distinct sensitivity to the initial state 
polarization selections. In some cases, particularly for the $(-++-)$ and 
$(+-+-)$ helicity choices, the differential cross section increases 
significantly for angles near $90^o$ taking on an m-like shape. This shape is, 
in fact, symptomatic of the spin-2 nature of the K-K graviton tower exchange 
since a spin-0 exchange leads only to a flattened distribution. 
Given the very large statistics available with a typical integrated luminosity 
of 100 $fb^{-1}$, even after angular cuts are applied to remove the forward 
and backward SM poles, it is clear that the $\gamma \gamma \to W^+W^-$ 
differential cross 
section is quite sensitive to $M_s$ especially for the two initial state 
helicities specified above.

\vspace*{-0.5cm}
\nn
\begin{figure}[htbp]
\centerline{
\psfig{figure=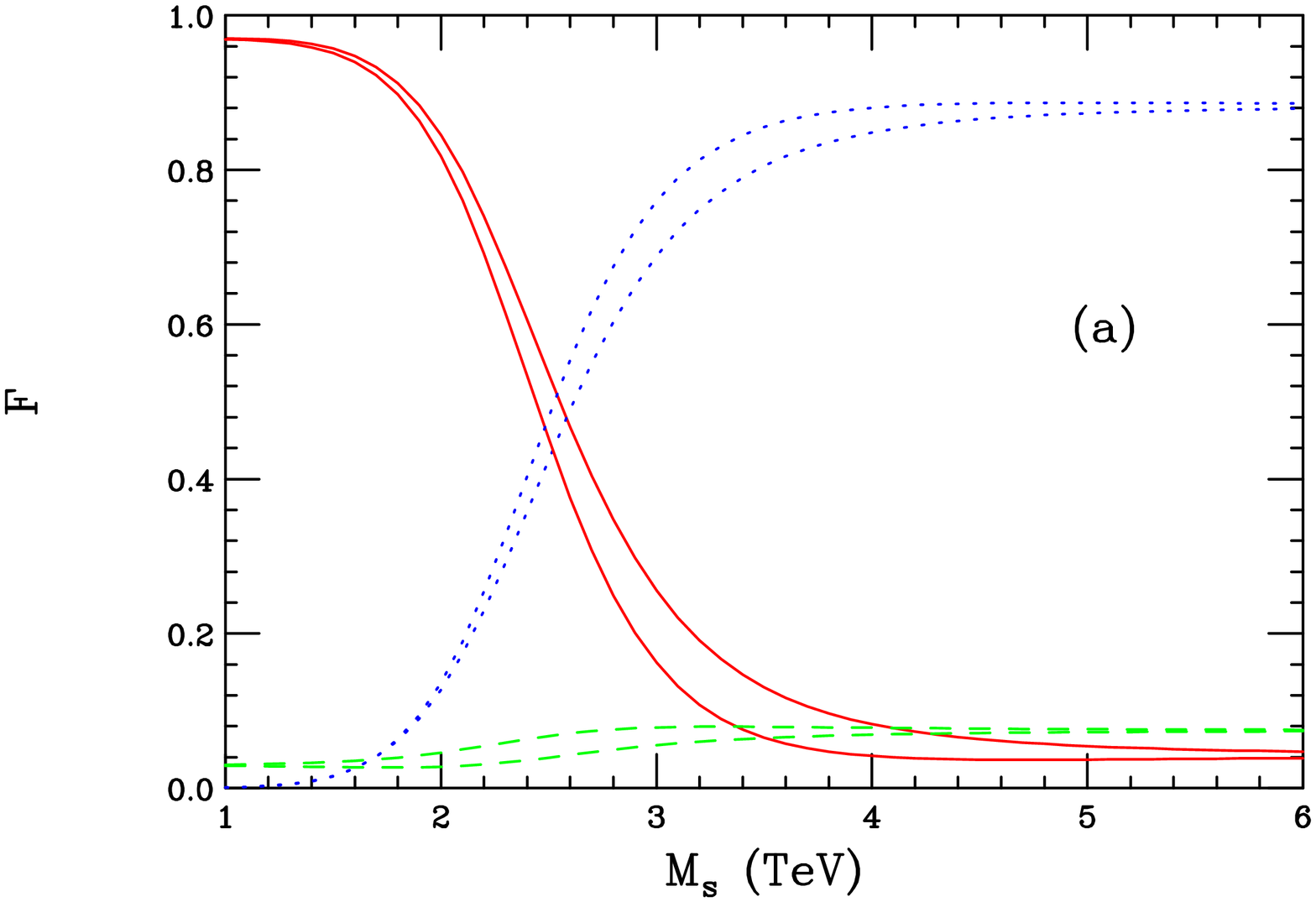,height=9.0cm,width=11.5cm,angle=90}}
\vspace*{5mm}
\centerline{
\psfig{figure=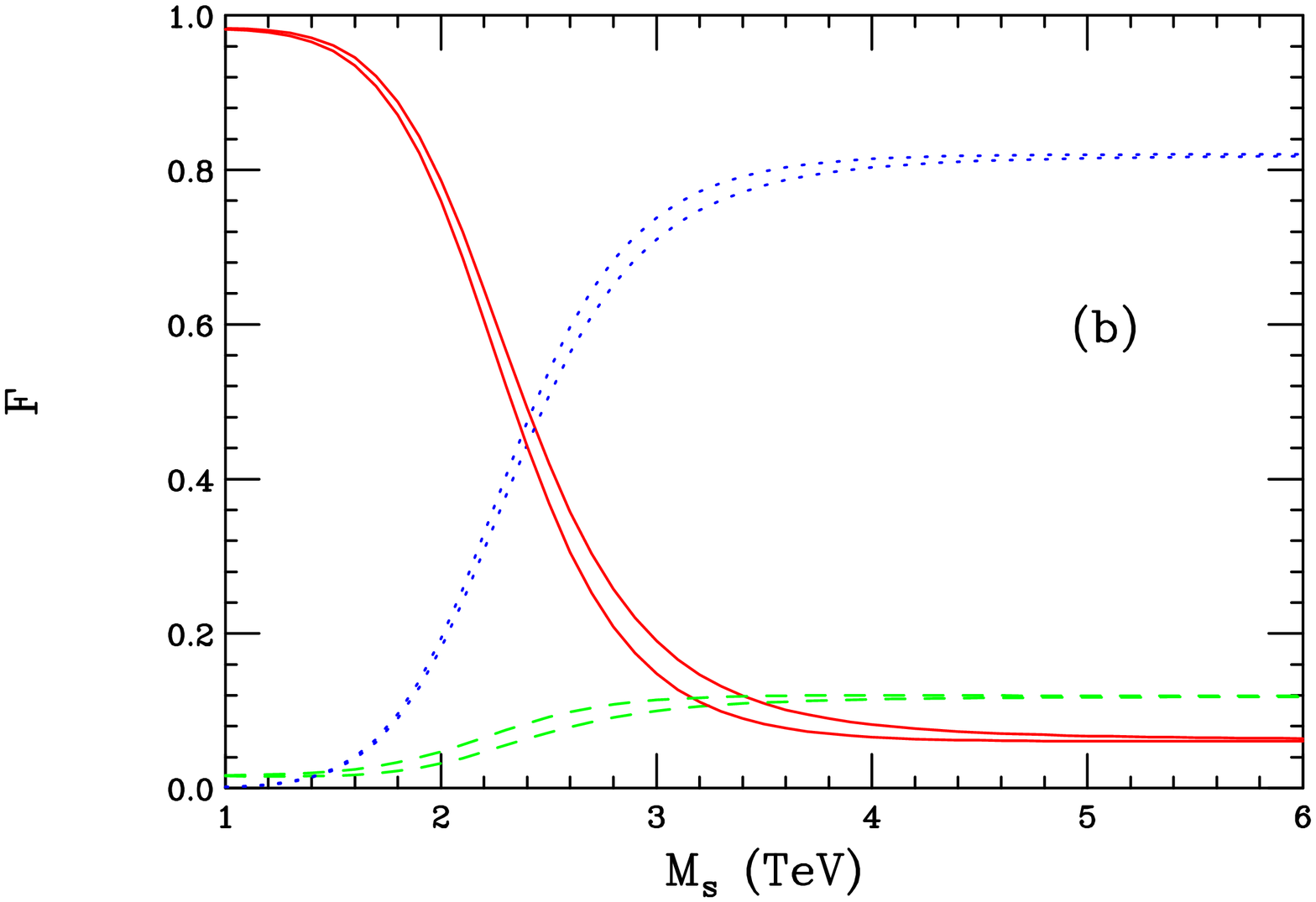,height=9.0cm,width=11.5cm,angle=90}}
\vspace*{0.1cm}
\caption{Fraction of LL(solid), TL+LT(dashed) and TT(dotted) $W^+W^-$ final 
states after angular cuts for the process $\gamma \gamma \to W^+W^-$ at a 
1 TeV $e^+e^-$ collider as a function of $M_s$ for either sign of $\lambda$. 
The 
initial state polarization in (a) is $(-++-)$ whereas in (b) it is $(+-+-)$.}
\label{fig2}
\end{figure}
\vspace*{0.4mm}

In addition to a significant modification to the angular distribution, the K-K
graviton tower exchange leads to another important effect through its 
influence on the polarizations of the two $W$'s in the final state. In the SM, 
independent of the initial electron and laser polarizations, the final state 
$W$'s are dominantly transversely polarized. Due to the nature of the spin-2 
graviton exchange, the K-K tower leads to a final state where both $W$'s are 
completely 
longitudinally polarized. Thus we might expect that a measurement of the 
$W$ polarization will probe $M_s$. To see this, we show in Fig.\ref{fig2}} 
the polarization fractions of the two $W$'s as a function of $M_s$ at a 1 TeV 
collider assuming two different choices of the initial state polarizations. 
In the results presented in this figure, an angular cut of $|z|<0.8$ has been 
applied to remove the SM poles in the forward and backward directions. 
Here we see that the fraction of final states where both $W$'s are 
longitudinal, denoted by $LL$, starts out near unity but falls significantly 
in the $M_s=2.5-3$ TeV region giving essentially the SM results above about 
5.5-6 TeV. The reverse situation is observed for the case where both $W$'s are 
transversely polarized, denoted by $TT$. The mixed case, denoted by the sum 
$TL+LT$, is also seen to grow from near zero to a modest value as $M_s$ 
increases and the SM limit is reached. Clearly a measurement of these final 
$W$ polarizations will allow us to probe respectable values of $M_s$. In 
order to determine the polarization fractions of the final state $W$'s one 
needs to examine the correlations in the 
angular distributions of the fermion decay products relative to 
the $W$'s original direction. If $\chi$ is the angle of one of the 
fermions with respect to the $W$ direction in the $W$ rest frame then 
transverse(longitudinal) $W$'s 
lead to an angular distribution $\sim 1 +(-) \cos^2 \chi$. Thus by measuring 
the correlation for both $W$'s the relevant $LL,TT$ and $LT+TL$ 
fractions can be extracted. In our numerical exercise we will assume 
that this can be done 
with an efficiency of $\sim 50\%$ for the $WW$ all hadronic final states ($W$ 
mass reconstruction removing combinatoric problems)
as well as for the $q\bar q\ell \nu$ final state with no significant 
backgrounds present 
in either case due to the very large $\gamma \gamma \to W^+W^-$ cross 
section.

\vspace*{-0.5cm}
\nn
\begin{figure}[htbp]
\centerline{
\psfig{figure=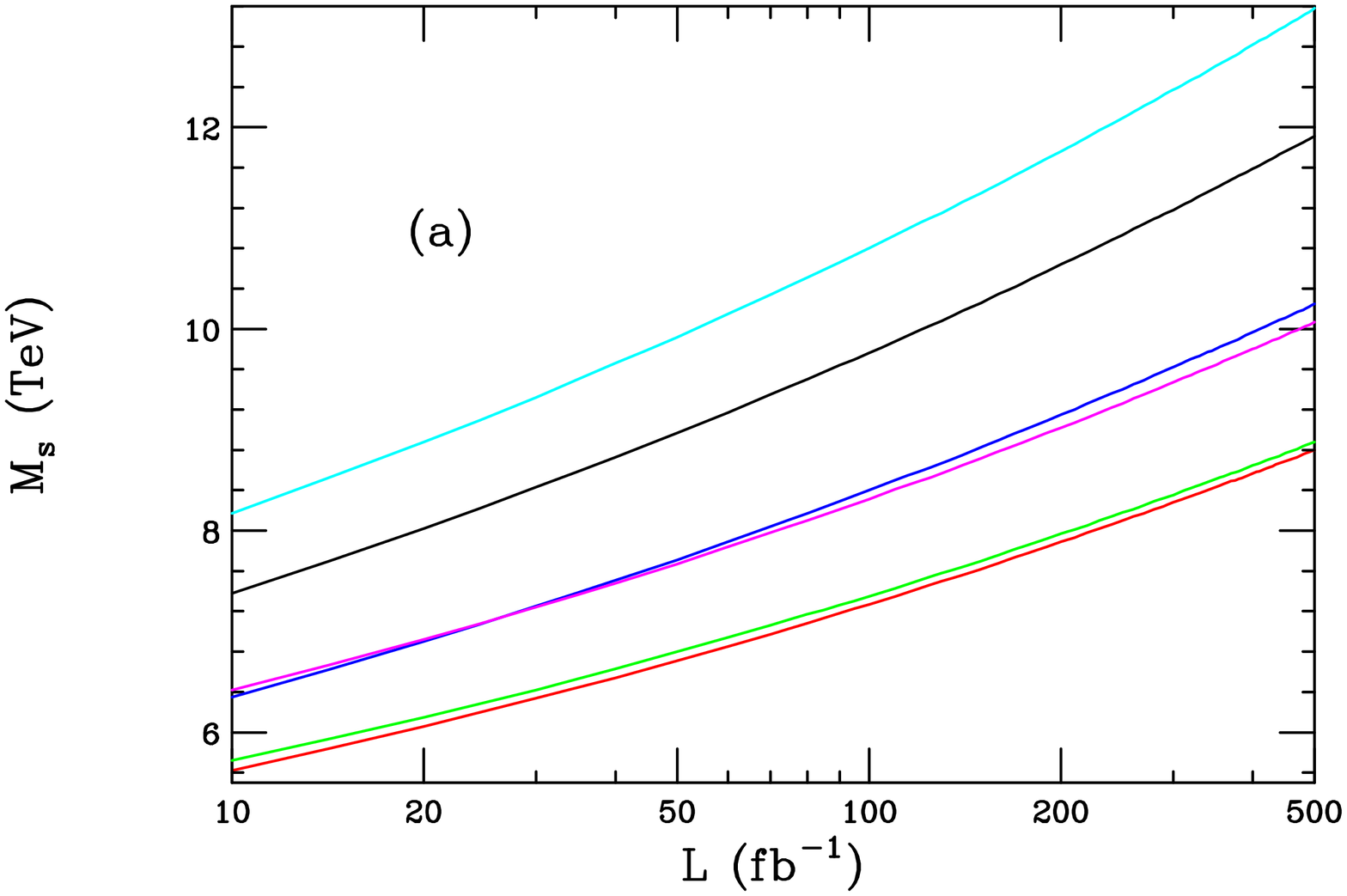,height=9.0cm,width=11.5cm,angle=90}}
\vspace*{5mm}
\centerline{
\psfig{figure=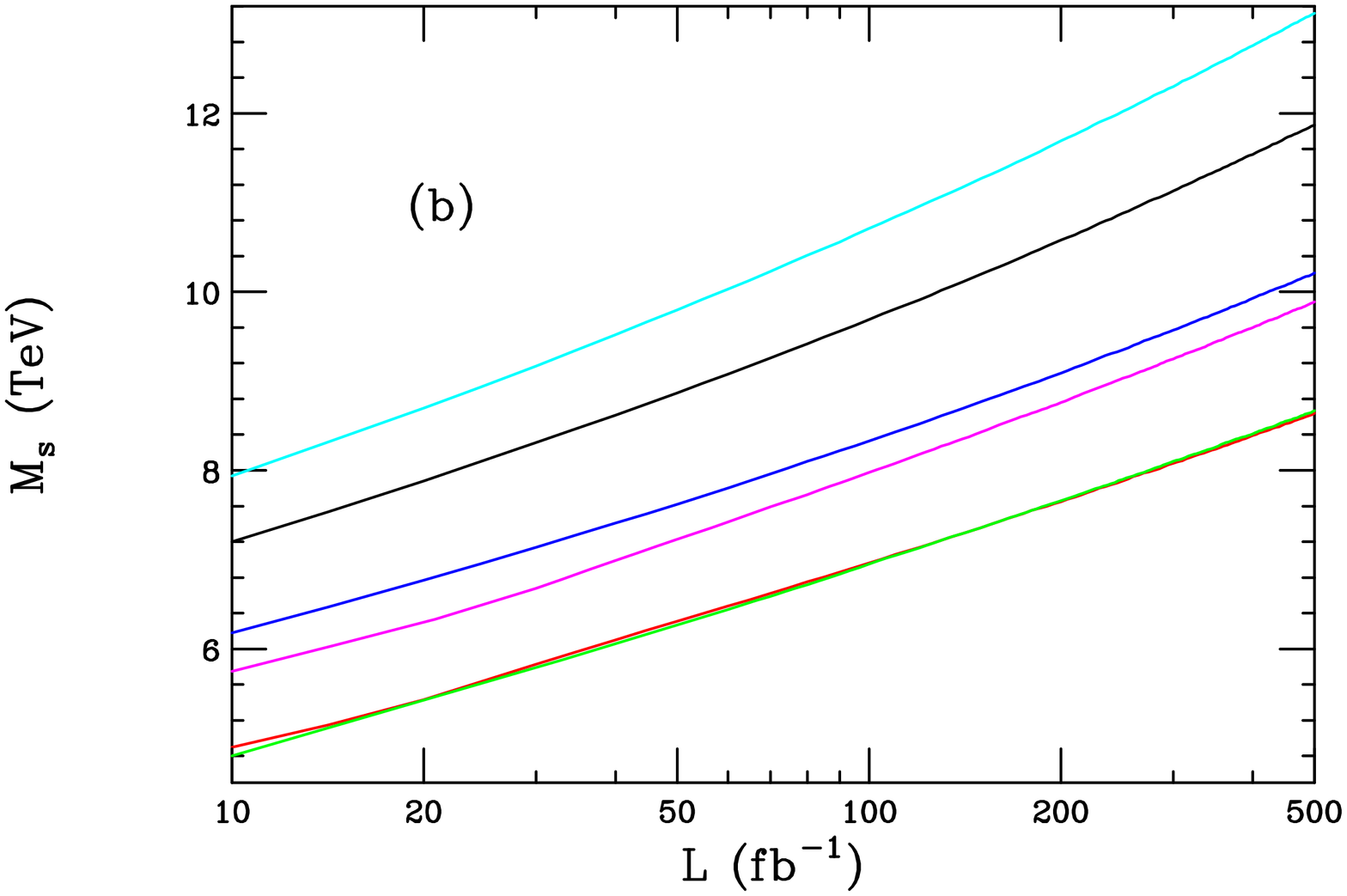,height=9.0cm,width=11.5cm,angle=90}}
\vspace*{0.1cm}
\caption{$M_s$ discovery reach from the 
process $\gamma \gamma \to W^+W^-$ at a 1 TeV $e^+e^-$ collider 
as a function of the integrated luminosity 
for the different initial state polarizations assuming (a)$\lambda=1$ or 
(b)$\lambda=-1$. From top to bottom on the right hand side of the figure the 
polarizations are $(-++-)$, $(+---)$, $(++--)$, $(+-+-)$, $(+---)$,  
and $(++++)$.}
\label{fig3}
\end{figure}
\vspace*{0.4mm}

By combining a fit to the total cross sections and angular distributions as 
well as the $LL$ and $LT+TL$ helicity fractions for various initial state 
polarization choices we are able to discern the discovery as well as the 
$95\%$ CL exclusion reaches for 
$M_s$. Given the rather steep behaviour of the both the SM-gravity and 
pure gravity terms in the cross section with $M_s$, \ie, $M_s^{-4}$ and 
$M_s^{-8}$ respectively, we do not expect these two reaches to differ 
significantly. 
In performing this analysis, in addition to the above efficiencies and 
angular cuts, we have assumed an overall integrated luminosity uncertainty of 
$1\%$. The results of this analysis for $\lambda=\pm 1$ and the six possible 
initial state polarizations are displayed in Figs.\ref{fig3} and \ref{fig4} 
which show both reaches as functions of the total integrated luminosity. 
Note both the strong sensitivity of the 
two reaches to the initial electron and laser polarizations as well as the 
large values obtainable particularly for the $(-++-)$ choice. In this 
particular case with 100 $fb^{-1}$ of integrated luminosity the discovery 
reach is almost $11\sqrt s$ for either sign of $\lambda$, which is greater 
than any other K-K graviton 
exchange process so far examined{\cite {pheno}}. Clearly, a more detailed 
analysis of these reaches, including both radiative corrections and detector 
effects, is more than warranted.

\vspace*{-0.5cm}
\nn
\begin{figure}[htbp]
\centerline{
\psfig{figure=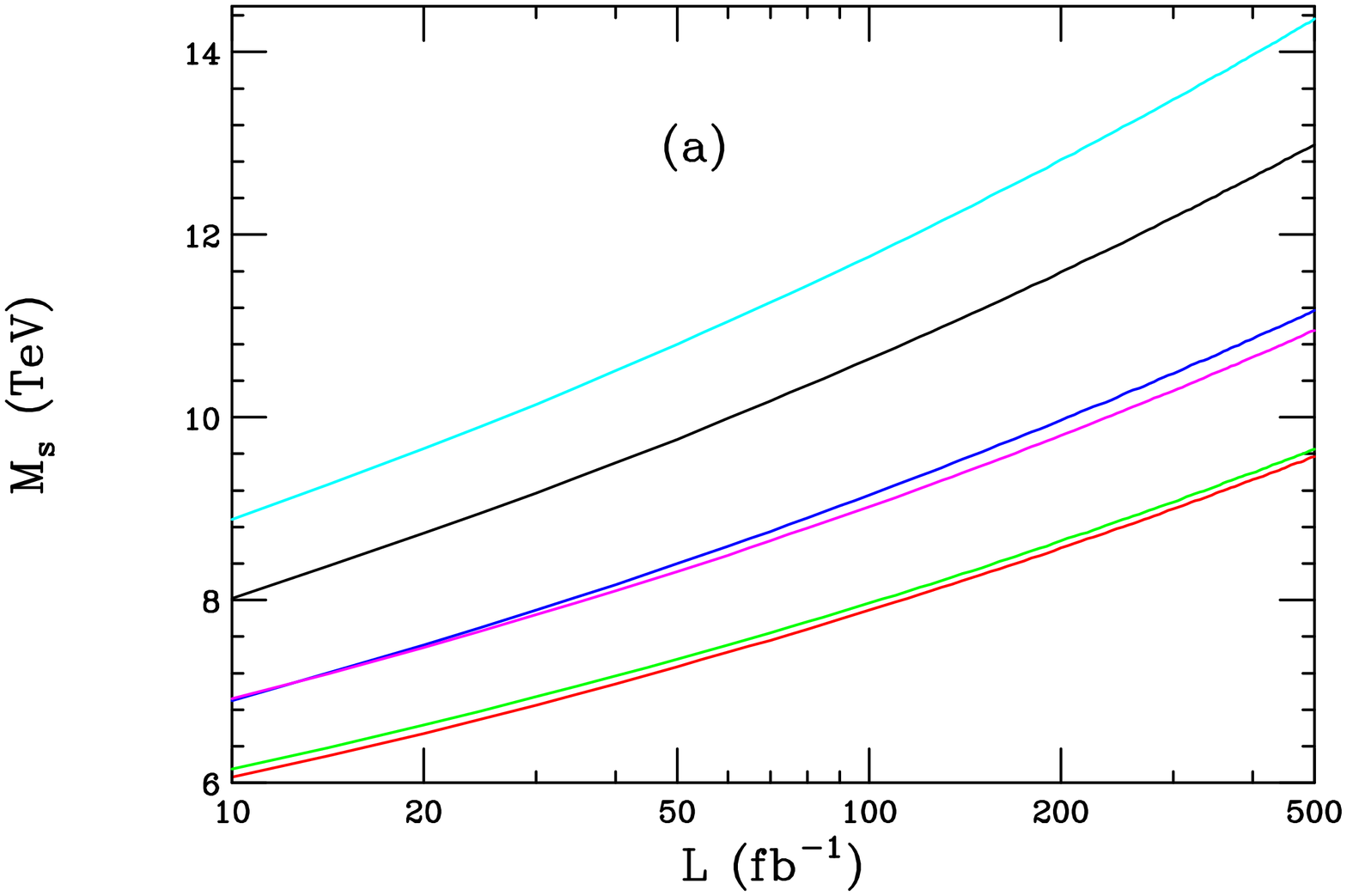,height=9.0cm,width=11.5cm,angle=90}}
\vspace*{5mm}
\centerline{
\psfig{figure=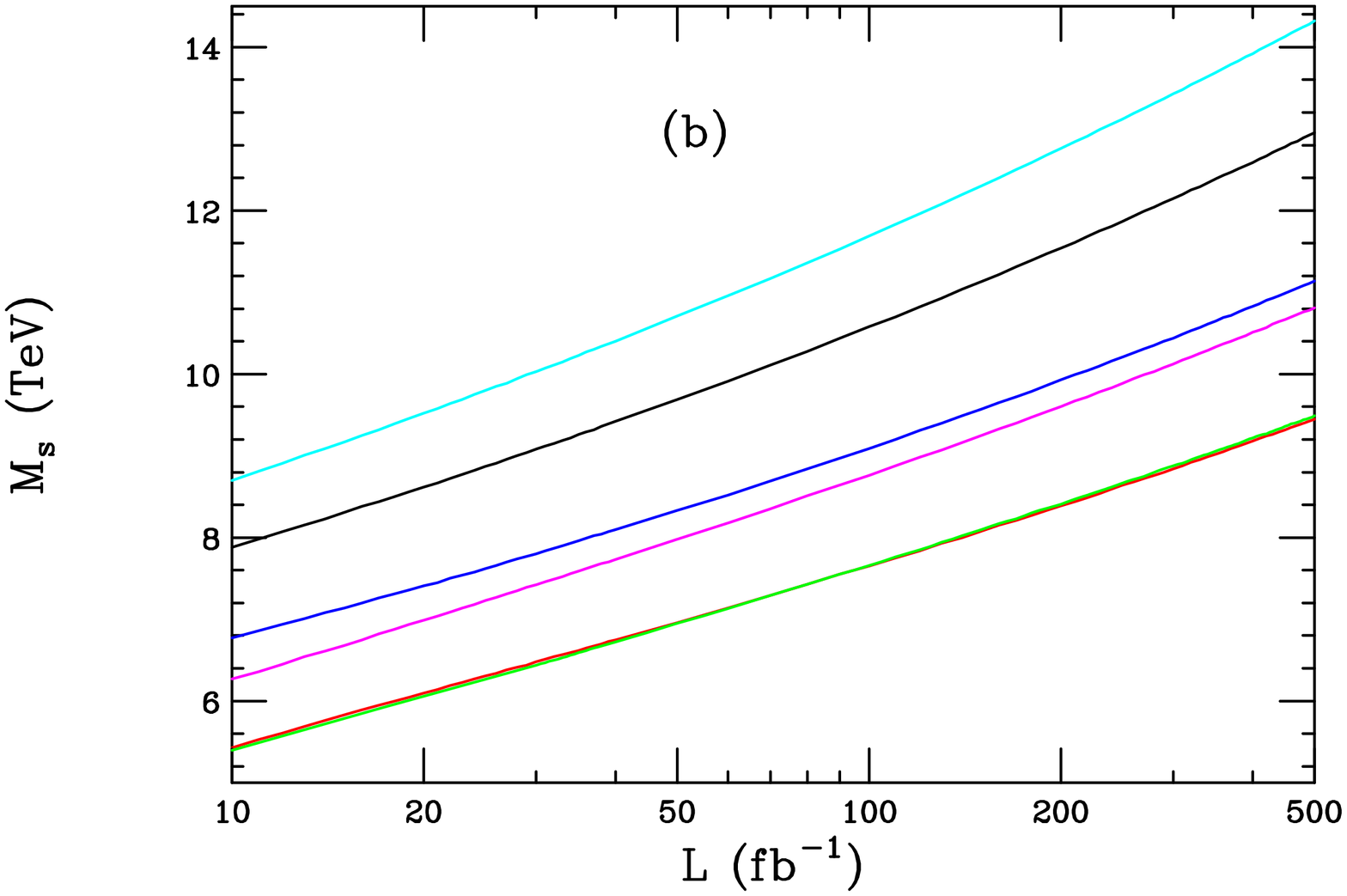,height=9.0cm,width=11.5cm,angle=90}}
\vspace*{0.1cm}
\caption{$95\%$ CL exclusion reach for $M_s$ at a 1 TeV $e^+e^-$ collider for 
the same cases as shown in the previous figure.}
\label{fig4}
\end{figure}
\vspace*{0.4mm}

There are of course many more observables available that are sensitive to the 
presence of the K-K graviton tower exchange and which can be included in a 
more detailed global fit; here we only mention two of them.
First, we can construct the invariant mass distribution of the two $W$'s in 
the final state approximately $90\%$ of the time and thus determine the 
differential cross section 
$d\sigma/ dM_{WW}$. After imposing the $|z|<0.8$ cut we would expect that 
the SM result will gradually fall off 
with increasing $M_{WW}$. However, since the K-K graviton tower exchange 
contribution grows quite rapidly 
with increasing $\hat s$, a modest value of $M_s$ will leads to 
an observable event excess at large values of $M_{WW}$. This is in fact 
exactly what we find as shown in Fig.\ref{fig5} for the sample choice of the 
initial state polarization $(-++-)$. Clearly a fit to this just distribution 
alone with the large statistics available will 
provide an additional probe of $M_s$ in the range beyond 4 TeV.

\vspace*{-0.5cm}
\nn
\begin{figure}[htbp]
\centerline{
\psfig{figure=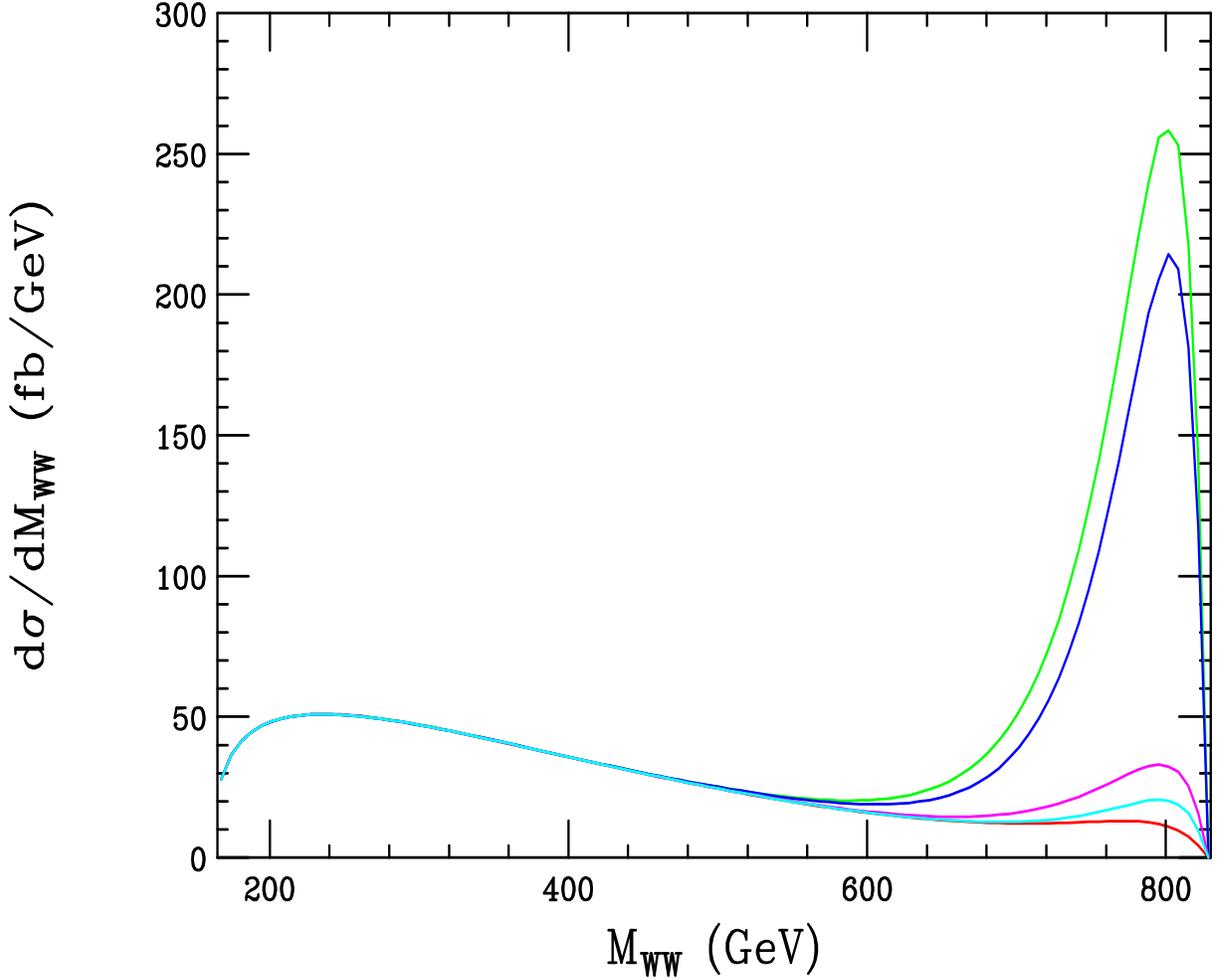,height=13cm,width=16cm,angle=90}}
\vspace*{0.1cm}
\caption[*]{$WW$ mass distribution for $\gamma \gamma \to W^+W^-$ at a 1 TeV 
$e^+e^-$ collider for the initial polarization of $(-++-)$ in the 
SM(red), the case of graviton exchange with $M_s$=2.5 TeV for both values of 
$\lambda$(green and blue, respectively) and for the corresponding scenario with 
$M_s$=3.5 TeV(magenta and cyan, respectively). In all cases a cut of $|z|<0.8$ 
has been imposed.}
\label{fig5}
\end{figure}
\vspace*{0.4mm}

Second, for the six possible initial state polarizations one can construct 
three independent polarization asymmetries of the form 
\begin{equation}
A_{pol}={{\sigma(P_{e1},P_{l1},P_{e2},P_{l2})-\sigma(P_{e1},P_{l1},-P_{e2},
-P_{l2})}\over {\sigma(P_{e1},P_{l1},P_{e2},P_{l2})+\sigma(P_{e1},P_{l1},
-P_{e2},-P_{l2})}}\,. 
\end{equation}
These asymmetries can also be made angular dependent, $A_{pol}(z)$, 
by interpreting the cross 
sections in both the numerator and denominator as differential in $z$ and 
taking advantage of the large statistics available. Further 
one may separately examine asymmetries constructed from distinct final state 
polarization states. Such a general class of asymmetries involving differences 
in initial state photon 
helicities form the basic ingredients of the Drell-Hearn-Gerasimov Sum 
Rule(DHG){\cite {dhg}} which has be applied, \eg , in the search for 
anomalous couplings between photons and $W,Z$'s{\cite {old}}. In order to 
prove the usefulness of these asymmetries let us first examine their angular 
dependencies in both the SM and when K-K towers of gravitons{\cite {gold}} 
are exchanged; 
this is shown in Fig.\ref{fig6} for a 1 TeV collider. Note that as 
$z\to \pm 1$ the SM dominates due to the large magnitude of the $u$- and 
$t$-channel poles. Away from the poles the three asymmetries all show a 
significant sensitivity to the K-K tower of graviton exchange. 
Although these 
asymmetries are not very big the large statistics of the data samples 
obtainable for this channel indicate that they will be very well determined
since many systematic errors will also cancel in forming the cross section 
ratios.

\vspace*{-0.5cm}
\nn
\begin{figure}[htbp]
\centerline{
\psfig{figure=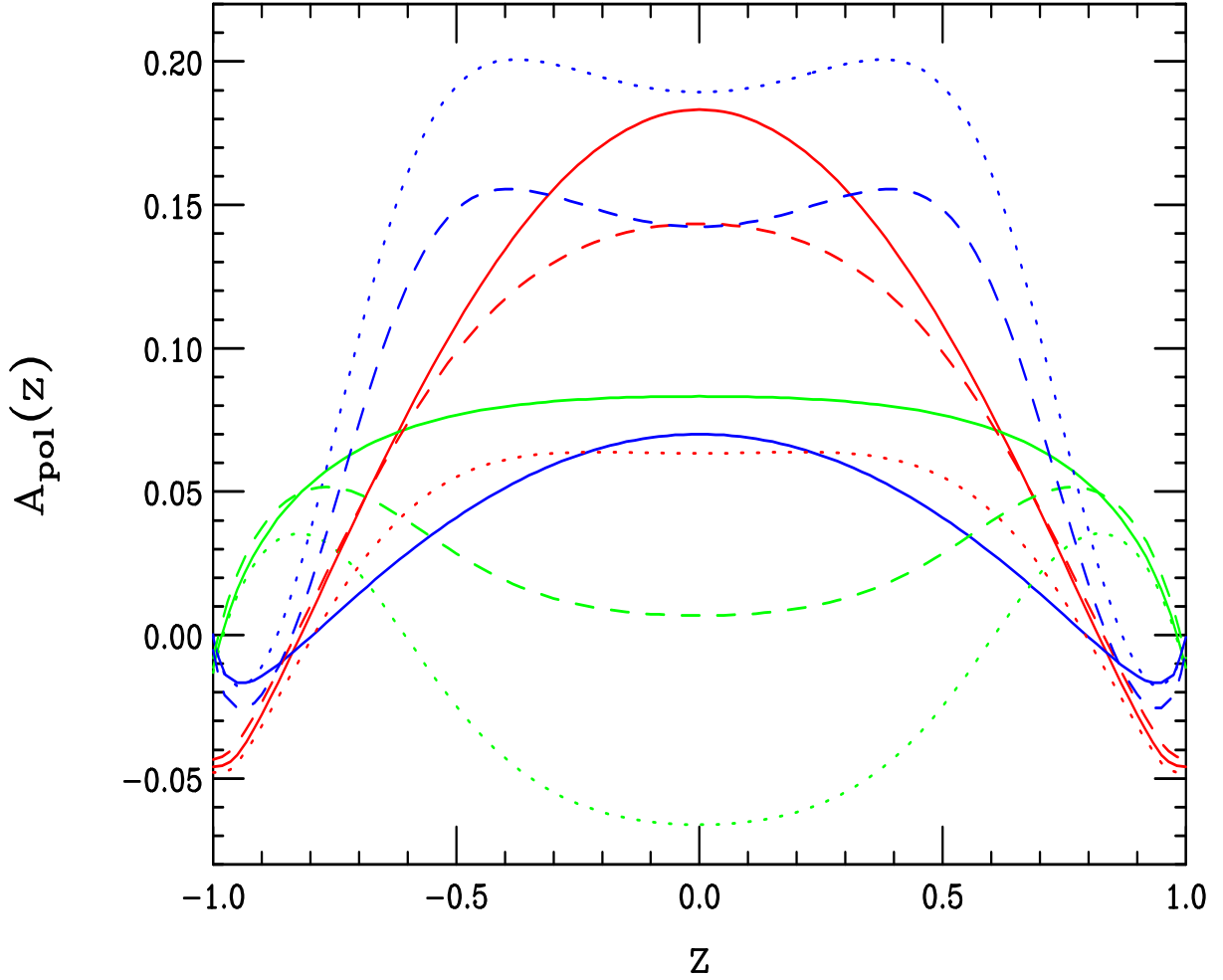,height=13cm,width=16cm,angle=90}}
\vspace*{0.1cm}
\caption[*]{Differential polarization asymmetries for 
$\gamma \gamma \to W^+W^-$ at a 1 TeV 
$e^+e^-$ collider for the SM(solid) as well with graviton tower exchange with 
$M_s$=2.5 TeV with $\lambda=\pm 1$(the dotted and dashed curves). We label the 
three cases shown by the first entry in the numerator in the definition of 
$A_{pol}$. Red represents an initial polarization of $(++++)$, green is for the 
choice $(+++-)$ and blue is for the case $(-++-)$.}
\label{fig6}
\end{figure}
\vspace*{0.4mm}

As is well known{\cite {old}} in order to satisfy the DHG sum rule 
the {\it integrated} asymmetry must pass through a zero (by the mean value 
theorem) for some value of $\hat s$. The exact position of the zero has been 
shown to be sensitive to new physics and to any applied kinematic cuts. What 
happens in the case where the SM is augmented by an exchange of a K-K tower 
of gravitons and how does $A_{pol}$ vary with $M_{WW}$? Fig.\ref{fig7} 
address these question directly for all three independent asymmetries after 
employing the $|z|<0.8$ cut. The position of the zero is seen to be the same 
in all three cases and quite insensitive to even low values of $M_s \sim 2.5$ 
TeV. This is due to the rather unfortunate fact that the zero occurs in the 
low $M_{WW}$ region where contributions from 
finite $M_s$ are very difficult to observe. 
The behaviour of $A_{pol}$ at larger values of $M_{WW}$ {\it is}, 
however, very sensitive to graviton exchange indicating a strong modification 
of the Sum Rule integral itself. Since the K-K graviton tower exchange is 
represented by an effective dimension-eight operator it can be shown for 
all initial state helicities that the integral in the Sum Rule diverges. This 
should be expected since the new operator is non-renormalizable.

\vspace*{-0.5cm}
\nn
\begin{figure}[htbp]
\centerline{
\psfig{figure=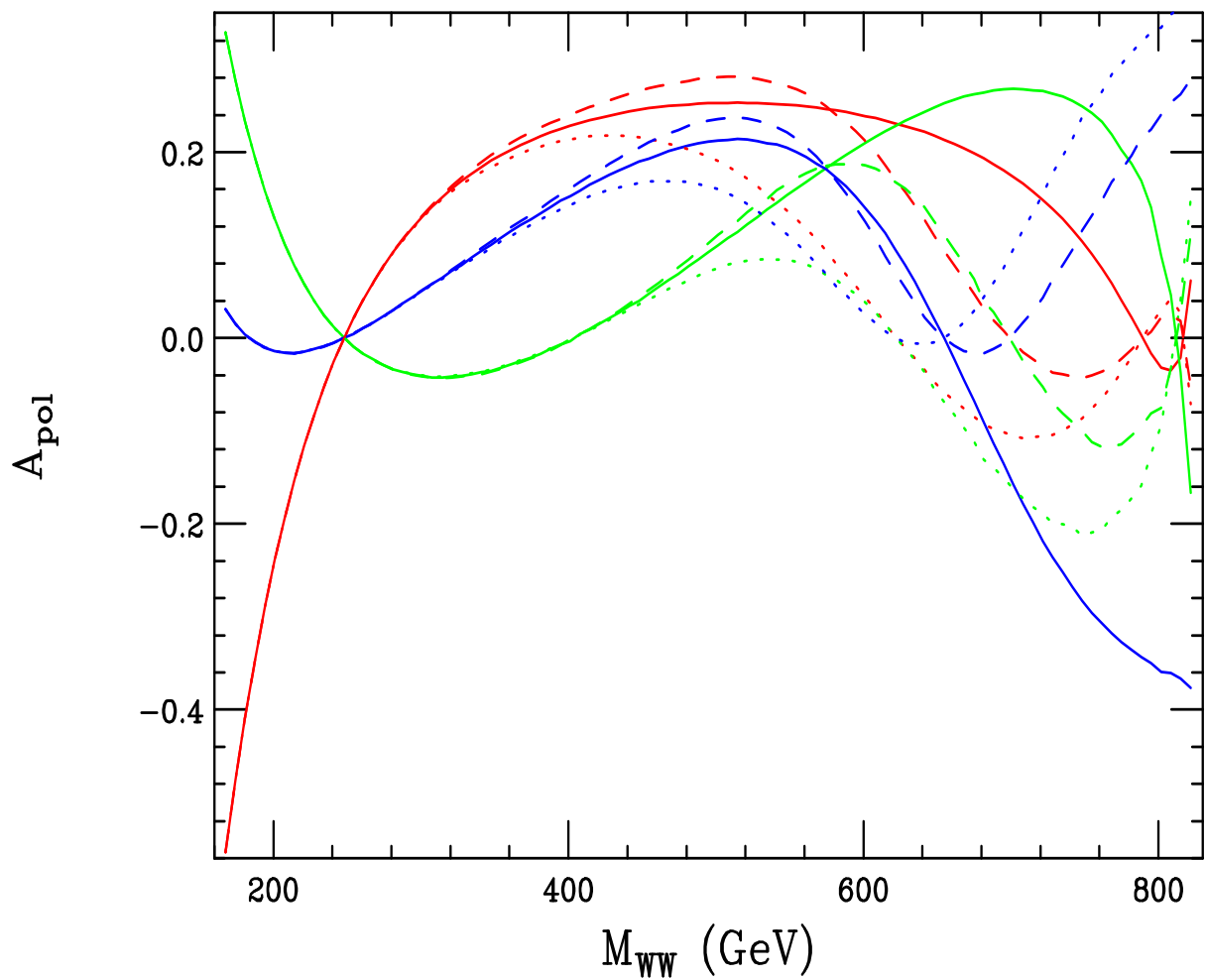,height=13cm,width=16cm,angle=90}}
\vspace*{0.1cm}
\caption[*]{Integrated polarization asymmetries for 
$\gamma \gamma \to W^+W^-$ at a 1 TeV $e^+e^-$ collider as functions of the 
$WW$ invariant mass. The labels for the various curves are as in the previous 
figure and a cut of $|z|<0.8$ has been applied.}
\label{fig7}
\end{figure}
\vspace*{0.4mm}

It is clear from the discussion above that there are a large number of 
observables that can be combined into a global fit to probe very high values 
of $M_s$ in comparison to the collider energy.
It should be noted however that due to the large statistics available 
the eventually determined discovery reach for $M_s$ using the 
$\gamma \gamma \to W^+W^-$ process will strongly depend on the size and 
variety of the experimental systematic errors.

\subsection{$\gamma \gamma \to ZZ$}

The process $\gamma \gamma \to ZZ$ does not occur at the tree level in the SM 
or MSSM. At the one loop level in the SM the dominant 
contribution{\cite {loop2}} arises from $W$ and 
fermion box diagrams and triangle graphs with $s$-channel Higgs boson 
exchange. (In SUSY models, additional contributions arise due to sfermion and 
gaugino loops as well as the other Higgs exchanges.) The $W$ bosons loops 
have been shown to be the dominant contribution with the fermions interfering 
destructively. This would seem to naively imply 
that this channel is particularly suitable for looking for new physics effects 
since the SM and MSSM rates will be so small due to the loop suppression. 
The SM cross section (which peaks in the forward and backward directions), 
after a cut of $|z|<0.8$, is found to be $\sim 80$ fb and almost purely 
transverse away from Higgs boson resonance peaks. The size of this cross 
section being only 
$\sim 10^{-3}$ of that for $W^+W^-$ will make the $ZZ$ final state difficult to 
find. Since it is not anticipated that the $W$ and $Z$ masses will be very 
well separated in the jet-jet channel at this level of rejection, we must 
demand that at least one of the 
$Z$'s decay leptonically reducing the effective number of useful $Z$ by a 
factor $\simeq 10$. (We can regain some reasonable fraction of this 
suppression depending 
on the practicality of the $q\bar q \nu \bar \nu$ final state. While this state 
is useful for constructing angular distributions assuming a very hermetic 
detector it cannot be used, \eg , to probe the polarization fractions of both 
of the $Z$'s. The addition of this final state for other analyses 
would yield an efficiency of $\sim 50\%$ instead of $\sim 10\%$.) 
A luminosity of 100 $fb^{-1}$ thus yields a sample of only 
about 800(4000) $Z$ pairs after these simple cuts and efficiencies are 
employed(assuming the $q\bar q \nu \bar \nu$ final states are included).

\vspace*{-0.5cm}
\nn
\begin{figure}[htbp]
\centerline{
\psfig{figure=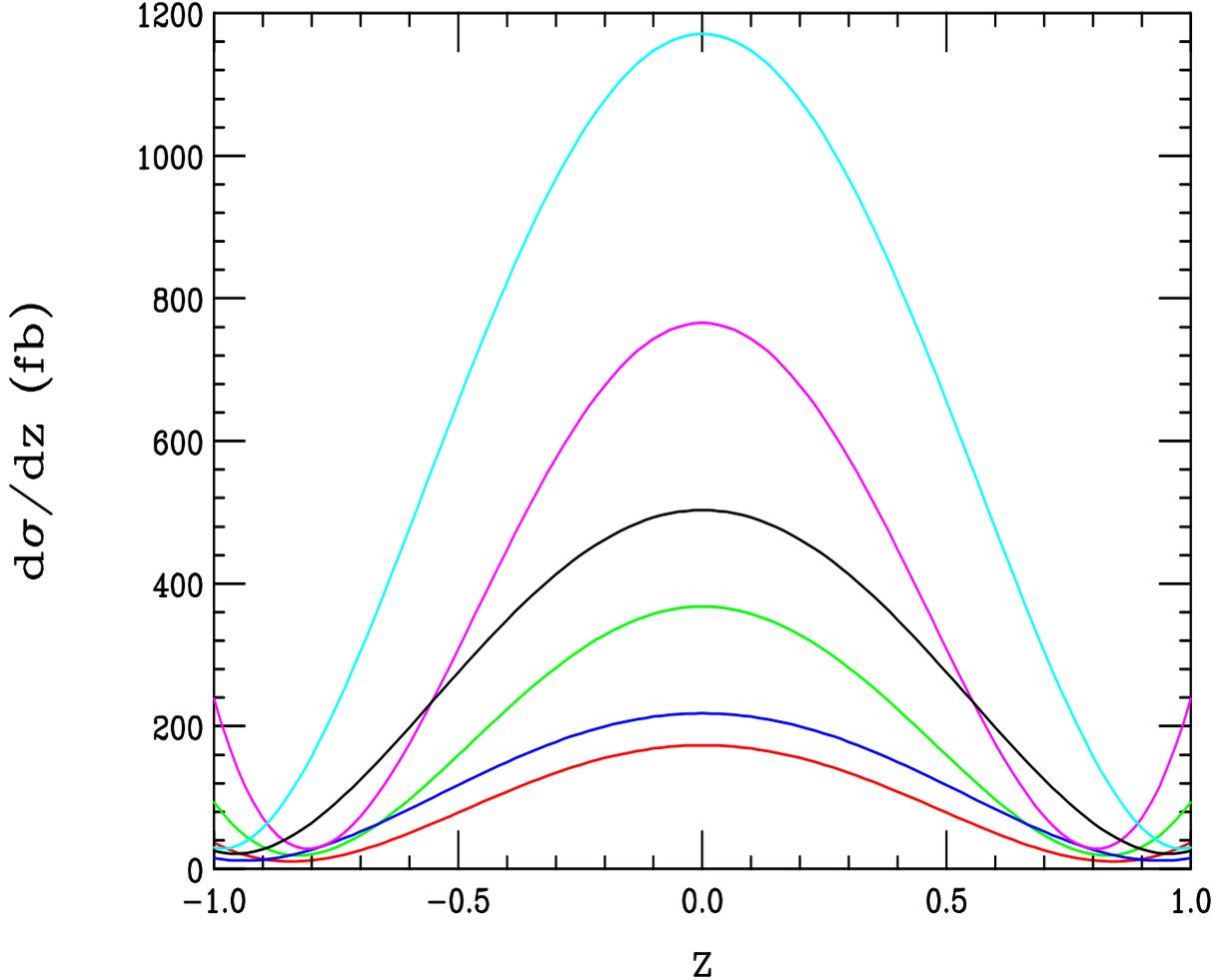,height=13cm,width=16cm,angle=90}}
\vspace*{0.1cm}
\caption[*]{Differential cross section for $\gamma \gamma \to ZZ$ at a 1 TeV 
$e^+e^-$ collider due to the exchange 
of a K-K tower of gravitons assuming $M_s=3$ 
TeV. From top to bottom in the center of the figure the initial state 
helicities are $(-++-)$, $(+-+-)$, $(+---)$, $(+++-)$, $(++--)$, $(++++)$.}
\label{fig8}
\end{figure}
\vspace*{0.4mm}

In the case of the ADD scenario the tree level K-K graviton tower contribution 
is now also present and is given by Eq.(2). Neglecting the loop-order SM 
contributions for the moment we obtain the resulting polarization-dependent 
differential 
cross sections shown in Fig.~\ref{fig8} where we have assumed $\sqrt s=1$ 
TeV and $M_s=3$ TeV for purposes of demonstration. Note that since this is 
the pure K-K graviton tower term there is no dependence here on the sign of 
$\lambda$. This cross section is found to scale with $s$ and $M_s$ as 
$\sim s^3/M_s^8$ and in contrast to the SM case is observed to peak at $90^o$. 
A short analysis along the lines of that performed 
above shows that essentially all of the $Z$'s in the final 
state are completely longitudinal with the $LL$ fraction being 
$\sim 99\%$ for the six possible initial state polarizations. Applying the 
same cuts, efficiencies and luminosities as above we find that for $M_s=4$ TeV 
this pure graviton contribution will only lead to 11(55) additional events 
beyond SM expectations which is not a huge excess. However, since the graviton 
tower cross section rises as $\hat s^3$ while the luminosities are falling off 
slowly, making a cut on the $ZZ$ invariant mass, 
$M_{ZZ}>550$ GeV, one finds that essentially all of the K-K-induced events lie 
above this cut (for the assumed value of the integrated luminosity) and would 
appear similar to 
a resonance-almost a broad Higgs-like{\cite {loop2}} bump. However, it would be 
doubtful that such an excess would be observed if $M_s>5$ TeV due to the 
$\sim M_s^{-8}$ scaling behaviour of the cross section unless significantly 
higher luminosities were available and systematic errors were very much under 
control. In addition, the use of the longitudinal polarization fraction would 
not seem to gain us any further reach. 

Although a detailed study of the loop-induced SM-graviton tower exchange 
interference terms have not yet been performed it is difficult to see how the 
search reach in this channel can exceed $\simeq 5$ TeV given the small 
magnitudes of the cross sections involved. Thus the anticipated overall reach 
for the $\gamma \gamma \to ZZ$ process is reasonably similar to that found 
for $\gamma \gamma \to \gamma \gamma$ {\cite {pheno}}.

\section{Summary and Conclusions}

$\gamma \gamma \to W^+W^-,ZZ$ offer new channels in which to search for the 
influence of graviton tower exchange, naively, each with their own individual 
strengths and weaknesses. As discussed above such channels are particularly 
clean for searches of the effects of graviton exchange since gauge boson towers 
cannot contribute to the cross sections at tree-level. 

We have found that 

\begin {itemize}

\item  The SM cross section for $\gamma \gamma \to W^+W^-$ is very large even 
after reasonable angular cuts are applied providing enormous statistics to 
look for new physics. This large cross section leads to an amplification of  
the size of the SM-graviton interference terms. The final state is quite 
clean there being little backgrounds due to rates alone and $90\%$ of the 
decay products are useable for analyses. The differential cross section 
as well as the 
polarization of the $W$'s in the final state were found to be quite sensitive 
to graviton exchanges especially for certain initial state electron and laser 
polarizations. The $W$-pair invariant mass distribution and the analogs of the
Drell-Hearn-Gerasimov polarization asymmetries were also shown to be able to 
probe large values of $M_s$. Fitting the cross section and final state 
polarizations, after cuts, efficiencies and systematic errors were included 
was shown to lead to search reaches $\sim 11\sqrt s$, the largest of any of 
the graviton exchange processes so far examined. Detailed simulations of this 
channel should be 
performed which include the other observables, radiative corrections and 
detector effects to verify or improve upon this reach estimate. 

\item  $\gamma \gamma \to ZZ$ would also appear to be an excellent process to 
probe for large values of $M_s$ since it only occurs at the loop level in the 
SM and MSSM. However both the SM and K-K tower cross sections are quite small 
for $M_s>4$ TeV even if the most advantageous initial state polarization 
choice, $(-++-)$, is assumed. Some assistance 
is gained by the fact that almost all 
of the $ZZ$ events will lie above a cut of $M_{ZZ}>550-600$ GeV and that they 
are almost purely longitudinally polarized. It seems unlikely, however, that 
the search reach in this channel can much exceed 5 TeV unless very large 
data samples become available.

\end{itemize}

The physics on large extra dimensions is quite exciting and may reveal itself 
at collider experiments in the not too far distant future.

\vskip1.0in

\noindent{\Large\bf Acknowledgements}

The author would like to thank J.L. Hewett, N. Arkani-Hamed, J. Wells, T. Han, 
J. Lykken, M. Schmaltz and H. Davoudiasl for multi-dimensional 
discussion related to this work. He would also like to thank M. Berger, 
D. Dicus and G.J. Gounaris for discussions related to the SM helicity 
amplitudes for $\gamma \gamma \to ZZ$.

\newpage

%
%%%%%%%%%%%%%%%%%%--- References
%%%%%%%%%%%%%%%%%%%%%%%%%%%%%%%%%%%%%%%%%%%%%%%%%%%%%%%
\def\MPL #1 #2 #3 {Mod. Phys. Lett. {\bf#1},\ #2 (#3)}
\def\NPB #1 #2 #3 {Nucl. Phys. {\bf#1},\ #2 (#3)}
\def\PLB #1 #2 #3 {Phys. Lett. {\bf#1},\ #2 (#3)}
\def\PR #1 #2 #3 {Phys. Rep. {\bf#1},\ #2 (#3)}
\def\PRD #1 #2 #3 {Phys. Rev. {\bf#1},\ #2 (#3)}
\def\PRL #1 #2 #3 {Phys. Rev. Lett. {\bf#1},\ #2 (#3)}
\def\RMP #1 #2 #3 {Rev. Mod. Phys. {\bf#1},\ #2 (#3)}
\def\NIM #1 #2 #3 {Nuc. Inst. Meth. {\bf#1},\ #2 (#3)}
\def\ZPC #1 #2 #3 {Z. Phys. {\bf#1},\ #2 (#3)}
\def\EJPC #1 #2 #3 {E. Phys. J. {\bf#1},\ #2 (#3)}
\def\IJMP #1 #2 #3 {Int. J. Mod. Phys. {\bf#1},\ #2 (#3)}

\end{document}